\documentclass[runningheads]{llncs}

\usepackage[T1]{fontenc}
\def\doi#1{\href{https://doi.org/\detokenize{#1}}{\url{https://doi.org/\detokenize{#1}}}}
\usepackage{graphicx}
\usepackage{listings}
\usepackage{amssymb}
\usepackage{amsmath}
\usepackage[misc]{ifsym}

\begin{document}
\title{Multi-scale Super-resolution Magnetic Resonance Spectroscopic Imaging with Adjustable Sharpness}
\titlerunning{Multi-scale Super-resolution MRSI with Adjustable Sharpness}
%
\author{Siyuan Dong\inst{1}\textsuperscript{(\Letter)} \and Gilbert Hangel\inst{2} \and Wolfgang Bogner\inst{2} \and Georg Widhalm\inst{3} \and Karl Rössler\inst{3} \and Siegfried Trattnig\inst{2} \and Chenyu You\inst{1} \and Robin de Graaf\inst{4} \and John Onofrey\inst{4} \and James Duncan\inst{1,4}}

%
\authorrunning{S. Dong et al.}
%
\institute{Electrical Engineering, Yale University, New Haven, CT, USA
\email{s.dong@yale.edu} \and Biomedical Imaging and Image-guided Therapy, Highfield
MR Center, Medical University of Vienna, Vienna, Austria \and Neurosurgery, Medical University of Vienna, Vienna, Austria \and Radiology and Biomedical Imaging, Yale University, New Haven, CT, USA}
\maketitle              
\begin{abstract}
Magnetic Resonance Spectroscopic Imaging (MRSI) is a valuable tool for studying metabolic activities in the human body, but the current applications are limited to low spatial resolutions. The existing deep learning-based MRSI super-resolution methods require training a separate network for each upscaling factor, which is time-consuming and memory inefficient. We tackle this multi-scale super-resolution problem using a Filter Scaling strategy that modulates the convolution filters based on the upscaling factor, such that a single network can be used for various upscaling factors. Observing that each metabolite has distinct spatial characteristics, we also modulate the network based on the specific metabolite. Furthermore, our network is conditioned on the weight of adversarial loss so that the perceptual sharpness of the super-resolved metabolic maps can be adjusted within a single network. We incorporate these network conditionings using a novel Multi-Conditional Module. The experiments were carried out on a \textsuperscript{1}H-MRSI dataset from 15 high-grade glioma patients. Results indicate that the proposed network achieves the best performance among several multi-scale super-resolution methods and can provide super-resolved metabolic maps with adjustable sharpness.

\end{abstract}
\section{Introduction}
Magnetic Resonance Spectroscopic Imaging (MRSI) is a non-invasive imaging technique that can provide spatial maps of metabolites in the tissue of interest. MRSI has become an invaluable tool in studying neurological diseases, cancer and diabetes \cite{de2019vivo}. Although hardware and acceleration techniques have continuously evolved \cite{bogner2021accelerated}, MRSI is limited to its coarse spatial resolutions due to the low concentration of metabolites. Therefore, a post-processing approach to increase the spatial resolution would greatly benefit MRSI applications. 

Most of the traditional post-processing methods for super-resolution MRSI rely on model-based regularization using anatomical MRI \cite{lam2014subspace,kasten2016magnetic,jain2017patch}, which often results in slow and unrealistic reconstructions. With the advances in deep learning-based image super-resolution techniques \cite{wang2020deep}, a few data-driven methods have been proposed for super-resolving MRSI metabolic maps and achieved promising results \cite{iqbal2019super,dong2021high}. These methods train neural networks to upscale a low resolution metabolic map to a higher resolution map under a fixed upscaling factor, for which we call single-scale super-resolution methods. However, the required upscaling factor depends on the MRSI application and is usually unknown before training the network. Training a separate network for each possible upscaling factor is time-consuming and requires large memory for storing network parameters. Additionally, due to the limited amount of in vivo MRSI data, the training dataset typically contains multiple metabolites, but the existing method treats all metabolites equivalently and does not consider the distinct spatial characteristics of each metabolite \cite{dong2021high}. Furthermore, the adversarial loss is incorporated to improve the perceptual quality of the super-resolved metabolic maps \cite{dong2021high}. It is well-known that more heavily weighted adversarial loss generates sharper images with more high-frequency details, whereas the risk of introducing artifacts (artificial features that make images look more realistic) also increases \cite{wang2020deep,kim2019constrained,dong2022invertible}. The current method requires training a separate network for each weight \cite{dong2021high} to tune the trade-off between image sharpness and image fidelity, which is again sub-optimal in terms of training time and memory. 

To tackle these limitations, we propose a unified MRSI super-resolution network built upon a novel Multi-Conditional Module (MCM) that can be conditioned on multiple hyperparameters: upscaling factor, type of metabolite and weight of adversarial loss. The proposed module efficiently integrates multiple functionalities into a single network, reducing the total training time and network size. This makes our method unique from previous works that only consider the upscaling factor \cite{wang2021learning,tan2020arbitrary}. Our network is able to (1) achieve comparable performance as the networks trained under a single-scale setting, (2) learn the super-resolution process based on the specific metabolite, and (3) provide multiple levels of sharpness for each super-resolved metabolic map.

\section{Methods}

\begin{figure}[t]
\centering
\includegraphics[width=0.95\textwidth, height=7.13cm]{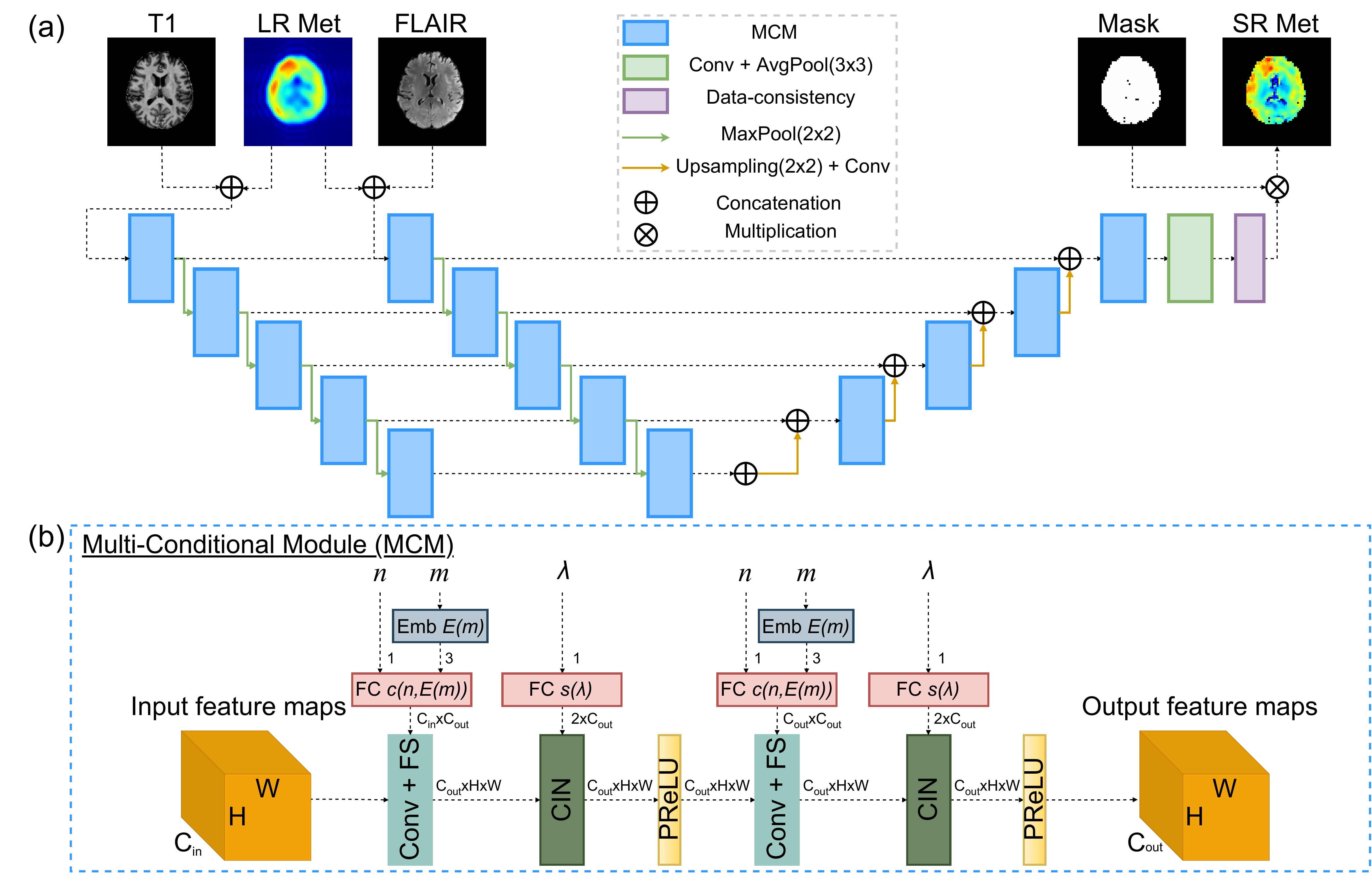}
\caption{Proposed network for multi-scale super-resolution MRSI with adjustable sharpness. (a) The overall architecture is a Multi-encoder UNet \cite{dolz2018dense}, which is appended with a Data-consistency module to guarantee that the corresponding k-space components of output images are consistent with experimental measurements \cite{schlemper2017deep}. We masked out the pixels in the network output that do not have valid ground truth due to quality-filtering. LR/SR Met: low-resolution/super-resolution metabolic maps. (b) The architecture of MCM. $n$, $m$ and $\lambda$ represent input resolution, type of metabolite and weight of adversarial loss. $C$, $H$ and $W$ are feature maps' channel number, height and width. The embedding layer $E(m)$ learns a transformation matrix that maps each metabolite name $m$ to a vector of length 3. This vector is concatenated with the the input resolution $n$ and fed into the fully-connected (FC) layers c(n, E(m)). Conv+FS: Convolution layer with Filter Scaling, Emb: embedding layer, CIN: Conditional Instance Normalization \cite{huang2017arbitrary}, PReLU: Parametric ReLU \cite{he2015delving}.} 
\label{fig1}
\end{figure}

Previous literature indicates that multi-parametric MRI contains meaningful spatial priors for super-resolution MRSI \cite{jain2017patch,lam2014subspace,iqbal2019super,dong2021high}. Hence, we provide our network with T1-weighted (T1) and fluid-attenuated inversion recovery (FLAIR) MRI in addition to the low resolution MRSI metabolic map as the input. We adopt the Multi-encoder UNet as the overall architecture (Fig. \ref{fig1}(a)) because it has been demonstrated to perform better than the single-encoder structure when processing information from multi-modal images \cite{dolz2018dense,dong2021high}. Our main innovation is incorporating multiple conditions into the network through a specialized MCM block (Fig. \ref{fig1}(b)), which is detailed in the following sections. 

\subsection{Multi-scale Super-Resolution}

Given that $L \in \mathbb{R}^{n\times n}$ is the low resolution metabolic map (we assume equal height and width $n$), the aim is to reconstruct a super-resolved map $S \in \mathbb{R}^{N\times N}$ using a neural network $F$: 
\begin{equation}
    S = F(L, \text{T1}, \text{FLAIR})
\end{equation}
such that $S$ is close to the high resolution ground truth $I \in \mathbb{R}^{N\times N}$. In real practices, the experimentally acquired metabolic map may have different matrix sizes depending on the acquisition protocol, i.e. $n$ is unknown. Assuming that the target matrix size $N$ is fixed, the upscaling factor $\frac{N}{n}$ is unknown before seeing the test data. To avoid training a separate single-scale network for each $n$, one straightforward approach is to train a network with a dataset mixed with all possible $n$. However, the network needs to learn a fixed set of parameters to compromise between different $n$, which could result in sub-optimal performance. One recently proposed blind super-resolution technique modulates the feature map statistics (mean and standard deviation) after the convolution layers under the guidance of degradation \cite{hui2021learning}. However, based on our experiments, only modulating the feature map statistics cannot fully exploit the modulation potential. Another work proposed to use auxiliary Hypernetworks \cite{ha2016hypernetworks} to predict the parameters of the main network under the guidance of upscaling factor \cite{hu2019meta}. While predicting all network parameters with Hypernetworks have great modulation capability, the amount of network parameters is significantly increased \cite{mok2021conditional}. 

To this end, we propose to use Filter Scaling \cite{alharbi2019latent,yang2021unified} to modulate the network parameters without significantly increasing the network size. Specifically, for a convolution layer with $C_{in}$ input channels and $C_{out}$ output channels, the convolution filters $f\in\mathbb{R}^{C_{in}\times C_{out}\times k \times k}$ ($k=3$) are scaled based on $n$: 
\begin{equation}
    f' = c(n)*f
\label{Equation_FS}
\end{equation}
where $f'$ is the scaled convolution filters. The function $c(n)\in\mathbb{R}^{C_{in}\times C_{out}}$ is implemented as fully-connected layers that output a scaling factor for each $k\times k$ filter. This Filter Scaling strategy mostly retains the modulation capability of Hypernetworks but avoids generating all convolution filters, which requires the output dimension of $c(n)$ to be $C_{in}\times C_{out} \times k \times k$. 

\subsection{Metabolite-awareness}
The main challenge of developing data-driven methods for enhancing MRSI is the lack of abundant in vivo data. For data augmentation, the existing method uses a dataset consisting of metabolic maps derived from multiple metabolites \cite{dong2021high}. However, different metabolites have distinct spatial characteristics, so the super-resolution process should not be identical. Based on this observation, we propose a metabolite-specific modulation of the network parameters. We use the same Filter Scaling strategy as in Eqn. \ref{Equation_FS} but with an additional input $E(m)$:
\begin{equation}
    f' = c(n, E(m))*f
    \label{Equation_FS2}
\end{equation}
where $m$ is the types of metabolite, such as glutamate (Glu) or glycine (Gly), and $E$ is a trainable embedding layer that converts words to numerical vectors. 

\subsection{Adjustable Sharpness}
Our loss function consists of pixelwise loss, structural loss and adversarial loss \cite{dong2021high}. The pixelwise loss penalizes the pixelwise difference between the network output $S$ and the ground truth $I$ using L1 norm, i.e. $L_{pixel} = \frac{1}{N^2}\sum^{N, N}_{i,j}|S_{i,j} - I_{i,j}|$. The structural loss maximizes a structurally-motivated image quality metric Multiscale Structural Similarity (MS-SSIM) \cite{wang2003multiscale,zhao2016loss}, i.e. $L_{structural} = 1-\text{MS-SSIM}(S, I)$. The adversarial loss \cite{wang2020deep} uses a discriminator (a 4-layer CNN followed by FC layers, trained alternatively with the generator) to minimize the perceptual difference between $S$ and $I$, which was implemented as a Wasserstein GAN \cite{gulrajani2017improved}. The overall loss is a weighted sum of the three components:

\begin{equation}
    Loss = (1-\alpha) L_{pixel} + \alpha L_{structural} + \lambda L_{adversarial}
    \label{Eqn6}
\end{equation}
where $\alpha\in [0, 1]$ and $\lambda\geq0$ are the hyperparameters. To be more specific, $\lambda$ controls a trade-off between image fidelity and image sharpness.  

We propose to condition our network on $\lambda$ such that this trade-off can be tuned within a single network. Inspired by a recent work on hyperparameter tuning \cite{mok2021conditional}, we use Conditional Instance Normalization \cite{huang2017arbitrary} to modulate the feature map statistics based on $\lambda$:

\begin{equation}
    y = s_{1:C}(\lambda)(\frac{x-\mu(x)}{\sigma(x)}) + s_{C+1:2C}(\lambda)
\end{equation}
where $\mu(x)\in\mathbb{R}^{C}$ and $\sigma(x)\in\mathbb{R}^{C}$ are the channel-wise mean and standard deviation of the feature map $x$ with channel number $C$. $s(\lambda)\in\mathbb{R}^{2C}$ is fully-connected layers that provide modulated channel-wise standard deviation in the first half of the output $s_{1:C}(\lambda)$ and modulated mean in the second half of the output $s_{C+1:2C}(\lambda)$. In this way, $\lambda$ controls the sharpness of the final output by modulating the feature maps after each convolution layer. 

\section{Experiments and Results}
\subsection{Data Acquisition and Preprocessing}
Due to the low concentration of metabolites, acquiring high resolution MRSI with acceptable SNR is always a challenge. We collected a unique 3D brain MRSI dataset from 15 high-grade glioma patients, with high resolution and high SNR \cite{hangel2020high}. \textsuperscript{1}H-MRSI, T1 and FLAIR were acquired with a Siemens Magnetom 7T whole-body-MRI scanner. Informed consent and IRB approval were obtained. The MRSI sequences were acquired in 15 min, with an acquisition delay of 1.3 ms and a repetition time of 450 ms. The measurement matrix is 64$\times$64$\times$39, corresponding to 3.4$\times$3.4$\times$3.4 mm$^3$ nominal resolution. The 3D metabolic maps were quantified from the voxel spectra using LCModel v6.3-1 \cite{provencher2014lcmodel}.
The spectra with insufficient quality (SNR $<$ 2.5 or FWHM $>$ 0.15 ppm) or close to the skull are rejected in a quality-filtering step. MRI images were skull-stripped and co-registered via FSL v5.0 \cite{smith2004advances}. 

From each 3D MRSI scan, we selected 11-18 axial slices that have sufficient voxels inside the brain, resulting in 2D metabolic maps of 64$\times$64. These maps were regarded as the high resolution ground truth ($N=64$). The corresponding low resolution maps were obtained by truncating $n\times n$ components at the center of k-space. We focus on 7 metabolites, namely total choline (tCho), total creatine (tCr), N-acetyl-aspartate (NAA), Gly, glutamine (Gln), Glu and inositol (Ins). The selected metabolites are important markers for the detection and characterization of tumors, stroke, multiple sclerosis and other disorders \cite{de2019vivo}. 

\subsection{Implementation Details}
The channel numbers for MCM are $8,16,32,64,128$ at each feature level. The numbers of fully-connected layers and latent feature sizes for $c(n)$, $c(n, E(m))$, $s(\lambda)$ are 5, 7, 5 and 32, 64, 64, respectively. Due to the limited number of patients, we adopted 5-fold cross-validation, and we used 9 patients for training, 3 for validation and 3 for testing in each fold. Therefore, the evaluations were performed on all 15 patients, which we believe makes our results reliable. The average numbers of metabolic maps used for training, validation and testing are 890, 297 and 297, respectively. The training data was substantially augmented using random flipping, rotation and shifting during training. $n$ is uniformly sampled from the even integers between 16 and 32, and $\lambda$ is uniformly sampled between 0 and 0.1. $\alpha$ is set to 0.84 as recommended in previous literature \cite{zhao2016loss}. All networks were trained with Adam optimizer \cite{kingma2014adam}, batch size of 8, initial learning rate of 0.0001 and 100 epochs. Experiments were implemented in PyTorch v1.1.0 on NVIDIA GTX 1080 and V100 GPUs, with a maximum memory usage of $\sim$2.3 GB. 

\subsection{Results and Discussion}
We first compare our Filter Scaling strategy with several other multi-scale super-resolution methods. In this part, the trainings were performed without the adversarial loss ($\lambda=0$), because adversarial training lacks a deterministic objective function, which makes it hard to reliably compare the learning capabilities of different networks \cite{salimans2016improved}. The second part demonstrates the functionality of adjustable sharpness, for which the adversarial loss was included in the training. 

\textbf{Multi-scale super-resolution and metabolite-awareness} To set a gold standard for the multi-scale super-resolution methods, we first trained 9 separate Multi-encoder UNet for $n=16,18,20,22,24, 26,28,30,32$ in a single-scale setting (denoted as Single-scale). As the baseline multi-scale super-resolution method, a Multi-encoder UNet was trained with $n$ randomly sampled from those 9 values but without any conditioning (denoted as Unconditioned). We also implemented the AMLayer that only modulates the features maps \cite{hui2021learning} and the Hypernetworks that predict all network parameters \cite{hu2019meta}, both conditioned on $n$. To ensure fairness, we set the same layer numbers and latent sizes for the Hypernetworks as for $c(n)$. Finally, we implemented two versions of our method, one uses Filter Scaling conditioned on $n$ (Eqn. \ref{Equation_FS}, denoted as Filter Scaling), and the other uses Filter Scaling conditioned on both $n$ and $m$ (Eqn. \ref{Equation_FS2}, denoted as Filter Scaling with Met). We performed the Wilcoxon signed-rank test on PSNR and SSIM for each pair of methods. As shown in Table \ref{tab1}, Filter Scaling achieves the same levels of PSNR and SSIM compared to the Single-scale networks (insignificant differences, $p$-value $>0.05$), whereas the other methods perform worse (significant differences, $p$-value $<0.05$). Note that Filter Scaling requires much fewer network parameters than HyperNetworks due to the smaller output dimension of fully-connected layers $c(n)$. With the incorporation of metabolite-specific modulation, Filter Scaling with Met achieves even higher metrics and performs the best among all the methods (significant differences, $p$-value $<0.05$).

\begin{table}[t]
\centering
\caption{Quantitative evaluations for different multi-scale super-resolution methods in terms of peak signal-to-noise ratio (PSNR), structural similarity index (SSIM), total training time (shown in GPU hours) and the total number of network parameters (shown in millions). PSNR and SSIM are presented in mean $\pm$ standard deviation calculated over all 5-fold cross validation and 9 values of $n$.}\label{tab1}
\begin{tabular}{|c|c|c|c|c|}
\hline 
Method & PSNR & SSIM & Train time & Params\\
\hline \hline 
Single-scale & 30.81 $\pm$ 2.90 & 0.9473 $\pm$ 0.0237 & 14.4h & 8.7M \\
\hline
Unconditioned & 30.65 $\pm$ 2.82 & 0.9458 $\pm$ 0.0244 & 1.6h & 1.0M  \\
AMLayer \cite{hui2021learning} & 30.70 $\pm$ 2.86 & 0.9462 $\pm$ 0.0242 & 1.9h & 1.5M \\
Hypernetworks \cite{hu2019meta} & 30.78 $\pm$ 2.88 & 0.9471 $\pm$ 0.0240 & 1.7h & 26M  \\
Filter Scaling (ours) & 30.81 $\pm$ 2.91 & 0.9474 $\pm$ 0.0238 & 1.7h & 3.9M \\
Filter Scaling with Met (ours) & \textbf{30.86 $\pm$ 2.92} & \textbf{0.9477 $\pm$ 0.0240} & 1.8h & 6.9M \\
\hline
\end{tabular}
\end{table}

\begin{figure}[t]
\centering
\includegraphics[width=\textwidth]{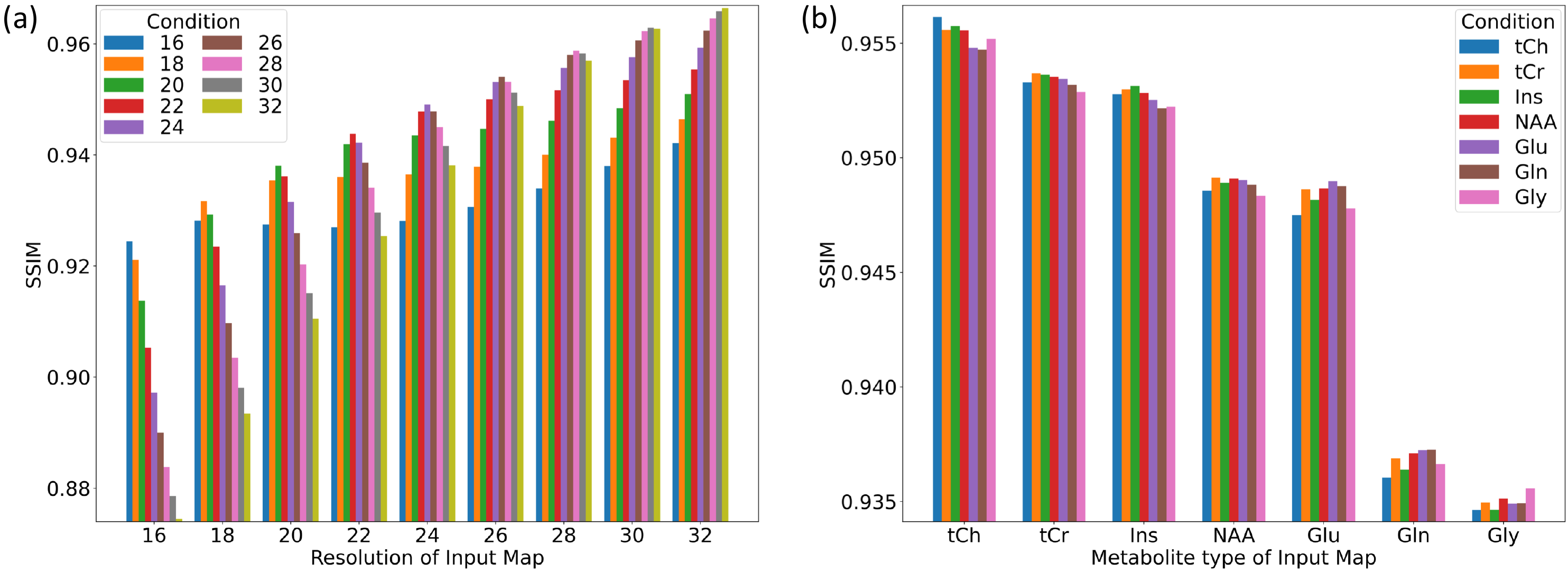}
\caption{Performance of our method under different combinations of network inputs (horizontal axis) and network conditions (color bars). (a) Study of the resolution. The plot shows the average SSIM computed over 5-fold cross-validation and all metabolites. (b) Study of the metabolite type. The plot shows the average SSIM computed over 5-fold cross-validation and all 9 values of $n$.} 
\label{fig2}
\end{figure}

To justify that our network learns the super-resolution process according to the conditions, we studied different combinations of input metabolic maps and conditions. The performance (measured using SSIM) is maximized when the resolution given in the condition matches with the input map's resolution (Fig. \ref{fig2}(a)), meaning that conditioning on $n$ helps the network to super-resolve the input map at the desired extent. Fig. \ref{fig2}(b) shows that the performance is maximized when the metabolite given in the condition matches with the metabolite of the input map (except only one failure case for NAA). This indicates that metabolite-specific modulation helps the network to super-resolve the input map based on each metabolite's distinct spatial characteristics.

\begin{figure}[t]
\centering
\includegraphics[width=0.95\textwidth]{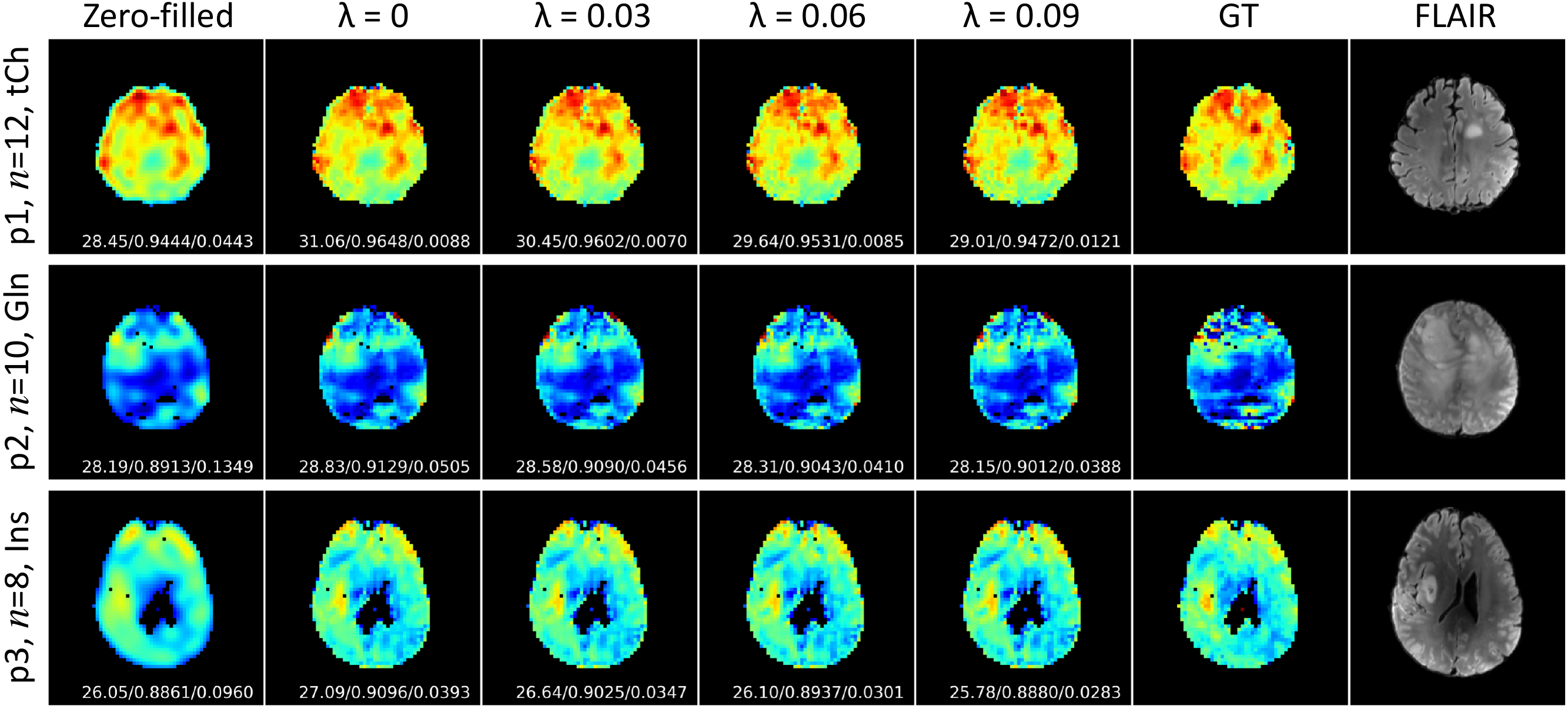}
\caption{Sharpness adjustability. From left to right are the standard k-space zero-filling, our method at $\lambda=0, 0.03, 0.06, 0.09$, ground truth (GT) and the FLAIR anatomical reference. From top to bottom are the super-resolution results of different metabolites tCh, Gln and Ins, each at a different input resolution ($n=12,10,8$) and from a different patient (p1, p2, p3). The numbers below each image are PSNR/SSIM/LPIPS.} 
\label{fig3}
\end{figure}

\begin{figure}[t]
\centering
\includegraphics[width=0.94\textwidth]{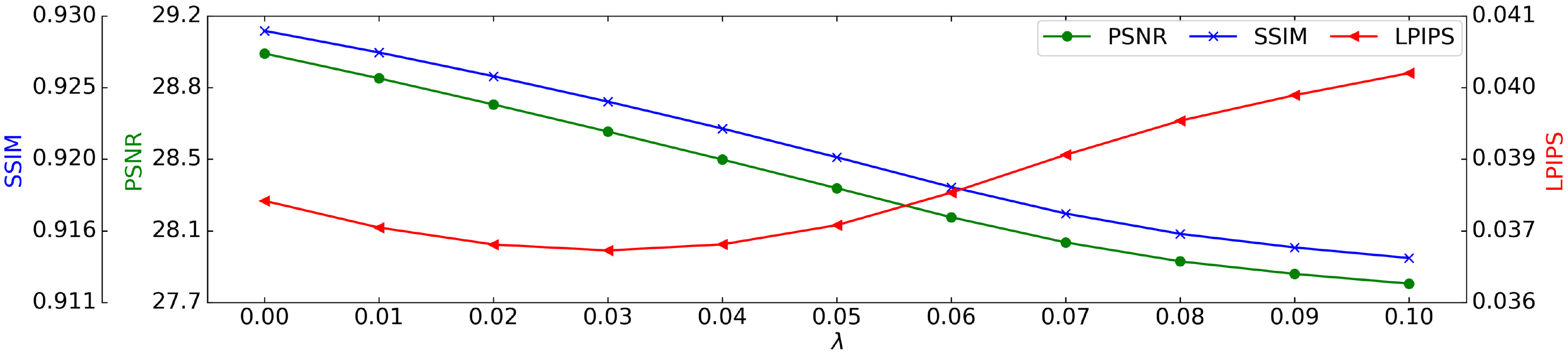}
\caption{Quantitative evaluations of different $\lambda$. The curves show mean scores of PSNR, SSIM and LPIPS, calculated over 5-fold cross validation and 9 values of $n$.} 
\label{fig4}
\end{figure}

\textbf{Adjustable Sharpness} Fig. \ref{fig3} shows a visualization of the super-resolved maps given by our network conditioning at $\lambda=0,0.03,0.06,0.09$, as well as the standard k-space zero-filling method. The results indicate that conditioning at $\lambda=0$ may generate metabolic maps that are blurry compared to the ground truth. Conditioning the network on greater levels of $\lambda$ can increase the image sharpness, especially at some high-frequency features. For example, in the last row of Fig. 3, the Ins hotspot at the tumor is blurred at $\lambda=0$. Using larger $\lambda$ gives better contrast at the tumor, which can be useful for better tumor detection in clinical studies. Besides, we report the quantitative analysis of the trade-off between image fidelity and image sharpness in Fig. \ref{fig4}. The image fidelity metrics, such as PSNR and SSIM, are inevitably degraded by the adversarial loss when a larger $\lambda$ is used and are therefore not sufficient to measure the improvement in perceptual quality. To measure the perceptual improvement quantitatively, we report the Learned Perceptual Image Patch Similarity (LPIPS), which was demonstrated to correlate well with human perceptual judgment \cite{zhang2018unreasonable}. The network achieves the best average LPIPS at around $\lambda=0.03$ (lower is better), meaning that the output maps are perceptually the most proximal to the ground truth at this level. Using $\lambda > 0.03$ results in worsened LPIPS, which means the images are at a high risk of being overly sharpened. Therefore, we recommend tuning the network between $\lambda=0$ and $0.03$, depending on how much image fidelity and image sharpness are pursued. 

\section{Conclusion}
This work presents a novel MCM to incorporate multiple conditions into a MRSI super-resolution network, which avoids training a separate network for each combination of hyperparameters. The network uses a Filter Scaling strategy to incorporate the upscaling factor and the type of metabolite into the learning process, outperforming several other multi-scale super-resolution methods. Moreover, our network is conditioned on the weight of adversarial loss, which provides sharpness adjustability. The method could potentially be applied clinically for faster data acquisition and more accurate diagnosis. For further validation, we can compare our results with tumor segmentation maps or histopathology to see if our method can help to better identify molecular markers for tumors. Future works will also extend the method to other MRSI applications, including mappings of other nuclei, e.g. \textsuperscript{2}H \cite{de2018deuterium,dong2020deep}, or other organs, e.g. liver \cite{coman2020extracellular}.

\subsubsection{Acknowledgements} This work was supported by the NIH grant R01EB025840, R01CA206180 and R01NS035193.

%
%
%
\bibliographystyle{splncs04}
\bibliography{reference}
%




\end{document}